\documentclass{article}

\usepackage{arxiv}

\usepackage[utf8]{inputenc} 
\usepackage[T1]{fontenc}    
\usepackage{hyperref}       
\usepackage{url}            
\usepackage{booktabs}       
\usepackage{amsfonts}       
\usepackage{nicefrac}       
\usepackage{microtype}      
\usepackage{lipsum}
\usepackage{graphicx}
\usepackage{amsmath}
\usepackage{caption}
\usepackage{textcomp}
\usepackage{bm}
\usepackage{subcaption}
\usepackage{mathtools, nccmath}
\graphicspath{ {./images/} }

\title{Sequence-to-Sequence Predictive Model: From Prosody To Communicative Gestures}

\author{
 Fajrian Yunus \\
  ISIR\\
  Sorbonne University\\
  Paris, France \\
  \texttt{fajrian.yunus@upmc.fr} \\
   \And
 Chlo\'e Clavel \\
  LTCI\\
  T\'el\'ecom-Paris, IP-Paris\\
  Paris, France \\
  \texttt{chloe.clavel@telecom-paristech.fr} \\
  \And
 Catherine Pelachaud \\
  ISIR\\
  CNRS / Sorbonne University\\
  Paris, France \\
  \texttt{catherine.pelachaud@upmc.fr} \\
}

\begin{document}
\maketitle
\begin{abstract}
Communicative gestures and speech acoustic are tightly linked. Our objective is to predict the timing of gestures according to the acoustic. That is, we want to predict when a certain gesture occurs. We develop a model based on a recurrent neural network with attention mechanism. The model is trained on a corpus of natural dyadic interaction where the speech acoustic and the gesture phases and types have been annotated. The input of the model is a sequence of speech acoustic and the output is a sequence of gesture classes. The classes we are using for the model output is based on a combination of gesture phases and gesture types. We use a sequence comparison technique to evaluate the model performance. We find that the model can predict better certain gesture classes than others. We also perform ablation studies which reveal that fundamental frequency is a relevant feature for gesture prediction task. In another sub-experiment, we find that including eyebrow movements as acting as beat gesture improves the performance. Besides, we also find that a model trained on the data of one given speaker also works for the other speaker of the same conversation. We also perform a subjective experiment to measure how respondents judge the naturalness, the time consistency, and the semantic consistency of the generated gesture timing of a virtual agent. Our respondents rate the output of our model favorably.
\end{abstract}


\section{Introduction} \label{sec:introduction}
Human naturally performs gestures while speaking~\cite{iverson1998people}. There are different types of communicative gestures which vary based  on the types of information they convey~\cite{mcneill1992hand} such as iconic (e.g., linked to the description of an object), metaphoric (e.g. conveying abstract idea), deictic (indicating a point in space) or beat (marking speech rhythm). Gesture helps the locutor to form what he or she wants to convey and also helps the listener to comprehend the speech~\cite{driskell2003effect}. Thus, it is desirable for a virtual agent which interacts with humans to show natural-looking gesturing behaviour. Because of that, researchers have been working on automatic gesture generation in the context of human-computer interaction~\cite{Cassell.et.al.01,kucherenko2019analyzing}. The techniques behind these generators are based on the principle that gestures and speech are related~\cite{mcneill1992hand}. Most of the prior gesture generators simplify the problem by focusing and generating only one type of gesture (e.g. beat inly or iconic only). There is also a recent work~\cite{kucherenko2020gesticulator} which tries to infer the gesture from both the speech acoustic and the text, which in principle enables the model to learn both the beat gestures and the semantic gestures. However, there is also a benefit of separating the learning of the gesture timing (when does it occur in relation to speech) from the learning of the gesture shape (the hands shape, wrist position, palm orientation, etc). By learning them separately, it would enable different models to be plugged in. On the other hand, if a model which does everything happens to not perform well on a certain task (e.g. generating the shape of semantic gestures), then fixing that weakness would require modifying the whole model.  In our current work, we first attempt to compute when a virtual-agent should perform a certain type of gesture. That is, we compute the gesture timing. We also simplify the problem by considering two categories of gesture: beat and other gesture types.

We compute the gesture class based on the speech acoustic. We learn their relationship by using a recurrent neural network with an attention mechanism~\cite{bahdanau2014neural}. The input is the sequence of speech prosody and the output is the sequence of gesture classes. Our input features are the fundamental frequency (F\textsubscript{0}), the F\textsubscript{0} direction score, and intensity. These features have been found to be highly correlated with gesture production. We also experiment with using other acoustic features that consider human perception of speech, namely the Mel-Frequency Cepstral Coefficients (MFCC) as the input features because they have been successfully used to generate body movements~\cite{hasegawa2018evaluation,kucherenko2019analyzing}. It should be noted that the model we are developing uses only the acoustic features as the input; the semantic feature is not considered yet. Our model aims to predict where gestures occur; more precisely the type of gestures (beat or ideational) and the timing of occurrence of gesture phases (stroke and other phases).  We are not yet dealing with the problem of predicting the form of the gestures nor which hand is used for the gesture. We will deal with this topic in a next step.

In Section \ref{sec:background} (Background), we explain the background concepts. In Section \ref{sec:related_work}, we explain the relevant prior works about gesture and gesture generation techniques. In Section \ref{sec:dataset}, we explain the dataset we use for our experiments. We explain the raw content and the various annotations provided in the dataset. In Section \ref{sec:feature_extraction}, we explain about how we extract usable data from the raw dataset. In Section \ref{sec:model}, we explain the model which we use and how it is implemented. In section \ref{sec:evaluation_measure}, we present the way we measure the performance of the model. In Section \ref{sec:experiment}, we describe our objective experiments. In Section \ref{sec:subjective_experiment}I, we describe our subjective experiment. In Section \ref{sec:discussion_and_conclusion}, we discuss our results and we draw the conclusions.  Finally, we explain our future direction in Section \ref{sec:future_work}.

\section{Background} \label{sec:background}
Gestures and speech are related. In most cases, communicative gestures only occur during speech~\cite{mcneill1992hand}. They are also co-expressive, which means that gestures and speech express the same or related meanings~\cite{mcneill1992hand}. They are also temporally aligned, that is gesture strokes happen at almost the same time as the equivalent speech segment~\cite{mcneill1992hand}. Gesture strokes themselves are known to occur slightly before or at the same time as the pitch accent~\cite{kendon1980gesticulation}. McNeill~\cite{mcneill1992hand} splits gestures into four classes, namely metaphorical, deictic, iconic, and beat. This classification is based on the information conveyed by the gesture. Metaphorical gestures are used to convey an abstract concept. Deictic gestures are used to point at an object or a location. Iconic gestures are used to describe a concrete object by its physical properties. Lastly, beat gesture does not convey any specific meaning, but it marks the speech rhythm.

The semantic gestures (communicative gestures other than beat, also called ``ideational gesture''~\cite{biancardi2017analyzing}) are characterized by temporal phases, namely preparation, pre-stroke-hold, stroke, post-stroke-hold, hold and retraction~\cite{kendon1980gesticulation}. The stroke phase carries the meaningful segment of a gesture; it is obligatory while the other phases are optional. Successive gestures may co-articulate one from the others. That is, when multiple gestures are performed consecutively, the gesture phases can be chained together. On the other hand, beat gestures do not have a phase~\cite{mcneill1992hand}. They are often produced with a soft open hand gestures and mark the speech rhythm.

Beat gestures can also be performed by facial and head movements~\cite{krahmer2007effects}. Specifically, it is noted that eyebrow movements can be related to beat gestures~\cite{krahmer2007effects}. It was observed that eyebrow movements tend to accompany prosodically prominent words~\cite{swerts2010visual}. It was also observed that pitch accents are accompanied by eyebrow movements~\cite{swerts2010visual,yasinnik2004timing,flecha2007non}.

\section{Related Work} \label{sec:related_work}
Embodied Conversational Agents (ECAs) are virtual agents endowed with the capacity to communicate verbally and non-verbally~\cite{Cassell.et.al.01}. We present existing works that aim to compute communicative gestures ECAs should display while speaking. Many researchers agree that gestures and speech are generated from a common process~\cite{mcneill1992hand,kendon1980gesticulation}. Most prior computational models simplify this relationship into that gestures can be inferred from speech. The earliest gesture generators for ECAs are rule-based~\cite{Cassell.et.al.01,lee2006nonverbal}. However, the relationship between speech and gestures is complex. Lately, to deal with the lack of precise knowledge, researchers develop machine-learning based gesture generators.

A common approach among the machine-learning based generators is generating a sequence of the gestures based on the acoustic. These techniques have a similar formulation: they express the problem as a time series prediction problem where the input is the acoustic and the output is the gesture motion. Hasegawa et al use Bi-Directional LSTM~\cite{hasegawa2018evaluation} with MFCC as their input. Kucherenko et al extend the work of Hasegawa et al by compacting the representation of the motion by using Denoising Autoencoder~\cite{kucherenko2019analyzing}. Kucherenko et al also experiment with other prosodic features, namely the energy of the speech signal, the fundamental frequency contour logarithm, and its derivative. Kucherenko et al report their technique yields a more natural movement. Ginosar et al~\cite{ginosar2019learning} use UNet and use MFCC as their input. They also add an adversarial learning component to enable mapping an input to multiple possible outputs. Ferst et al~\cite{ferstl2019multi} expand the use of adversarial learning further. They use multiple discriminators to evaluate the generated motion according to several qualities: phase structure, motion realism, intra-batch consistency, and displacement. They use fundamental frequency (F\textsubscript{0}) and MFCC as their input. Interestingly, embedded within their model architecture, there is a phase classifier. The classifier takes the three-dimensional velocities of the joints and the F\textsubscript{0} as input and yield the phase (preparation, hold, stroke, and other). The purpose of the phase classifier is to enforce of a realistic phase structure. For example, a preparation cannot be immediately followed by a retraction.

Among the machine-learning based generators, there are also text-based generators. Their aim is to generate ideational gestures. These gestures are related to the semantics, which are inferred from the text. Bergmann et al~\cite{bergmann2009gnetic} use Bayesian Decision Network to generate iconic gestures, by using the referent features and the pre-extracted discourse context. Ishii et al~\cite{ishii2018generating} use Conditional Random Field to generate a whole body pose. This technique does not model temporal dependency: the technique works at the level of phrase and the dependency between consecutive phrases is not modeled. Ahuja and Morency~\cite{ahuja2019language2pose} use a joint-embedding of text and body pose. The text is processed by using Word2Vec~\cite{mikolov2013distributed}. The technique generates whole body pose including arm movement.

There are also approaches which learns gesture statically. Nihei et al~\cite{nihei2019determining} use neural network to statically learn iconic gestures from a set of images. They use various images of similar objects, feed them into a pre-trained image-recognition neural network, and then extract the simplified shapes from the network. These simplified shapes can be reproduced as iconic gestures. L{\"u}cking et al~\cite{lucking2016finding} attempt to statically gather the typical metaphoric gesture for each image schema by running a human experiment. Image schema itself is a recurring pattern of reasoning to map one entity to another~\cite{johnson2013body}.

There is also an approach which use both the acoustic and the text. Kucherenko et al~\cite{kucherenko2020gesticulator} build a neural network model which takes both the text and the acoustic as the input to generate body movements. The text is represented as BERT embedding and the acoustic is represented by log-power mel-spectrogram. Because this technique takes both the text and acoustic as the input, in principle it can generate both the beat gestures and the ideational gestures. Interestingly, in their subjective study, they find that their respondents have a low agreement on which segments of the gestures represent the semantic.

Some interesting developments we observe in the recent work are the shift toward neural network~\cite{hasegawa2018evaluation,kucherenko2019analyzing,ginosar2019learning,ferstl2019multi,ahuja2019language2pose,ishii2020impact,nihei2019determining,kucherenko2020gesticulator}, the use of adversarial learning~\cite{ginosar2019learning,ferstl2019multi}, and the use of word embedding~\cite{ahuja2019language2pose,ishii2020impact,kucherenko2020gesticulator}. Neural network has been successful in recent years, which therefore makes it into a reasonable choice for machine learning problems. Adversarial learning enables one input to be correctly mapped to multiple different outputs. Effectively, it allows the same acoustic input to be mapped into different body movements. Word embedding is a representation of word as a vector. Two similar words will have their corresponding vectors also close to each other. Therefore, given 2 similar text inputs, the outputs would also be similar.

Our work is a bridge between the acoustic-based generators and the text-based generators. We attempt to tell when a virtual agent should perform a certain type of gesture. We distinguish beat gestures from ideational. First of all, the ideational gestures convey a specific meaning, beat gestures mark the speech rhythm. Moreover, beat gestures tend to appear during the theme while ideational gestures tend to appear \cite{Cassell.et.al.01} during the rheme that carry the new information~\cite{halliday1973explorations}. Additionally, we also distinguish the stroke phase from the other phases because the stroke phase is known to usually be near the pitch accent~\cite{kendon1980gesticulation}. Although it can be argued that a technique which learns the body movements from both the text and the acoustic also implicitly  learns the timing, there is also a benefit from separating the learning of the timing and the shape. By learning them separately, it would enable different models to be plugged in. For example, as Kucherenko et al~\cite{kucherenko2020gesticulator} find in their subjective study, the respondents have a low agreement on which segments are actually the semantic gestures. It suggests that the semantic gestures they generate probably happen to be not so prominent. However, because they have only one model, attempts to make the semantic gestures more prominent might change something else. Our usage of gesture phases as the classes is similar to \cite{ferstl2019multi}. However, their phases do not differentiate between beat and ideational gestures. Besides that, their phase classifier takes both acoustic and body movements as input while we do not have any body movement data. We evaluate our results by using a sequence comparison technique which tolerates shift and dilation. The spirit is similar to adversarial learning: for each input, there can be multiple correct output.

\section{Dataset} \label{sec:dataset}
We use the Gest-IS English corpus~\cite{saintamand2018gestis}. The corpus consists of 9 dialogues of a dyad, a man and a woman, discussing various topics in English face to face. The total duration is around 50 minutes. In those dialogues, the speakers are talking about physical description of some places, physical description of some people, scenes of two-person interactions, and instructions to assemble a wooden toy.

\vspace{0.00mm}

\begin{figure}[!htb]
\begin{minipage}{1.0\linewidth}
  \centering
  \centerline{\includegraphics[width=6cm]{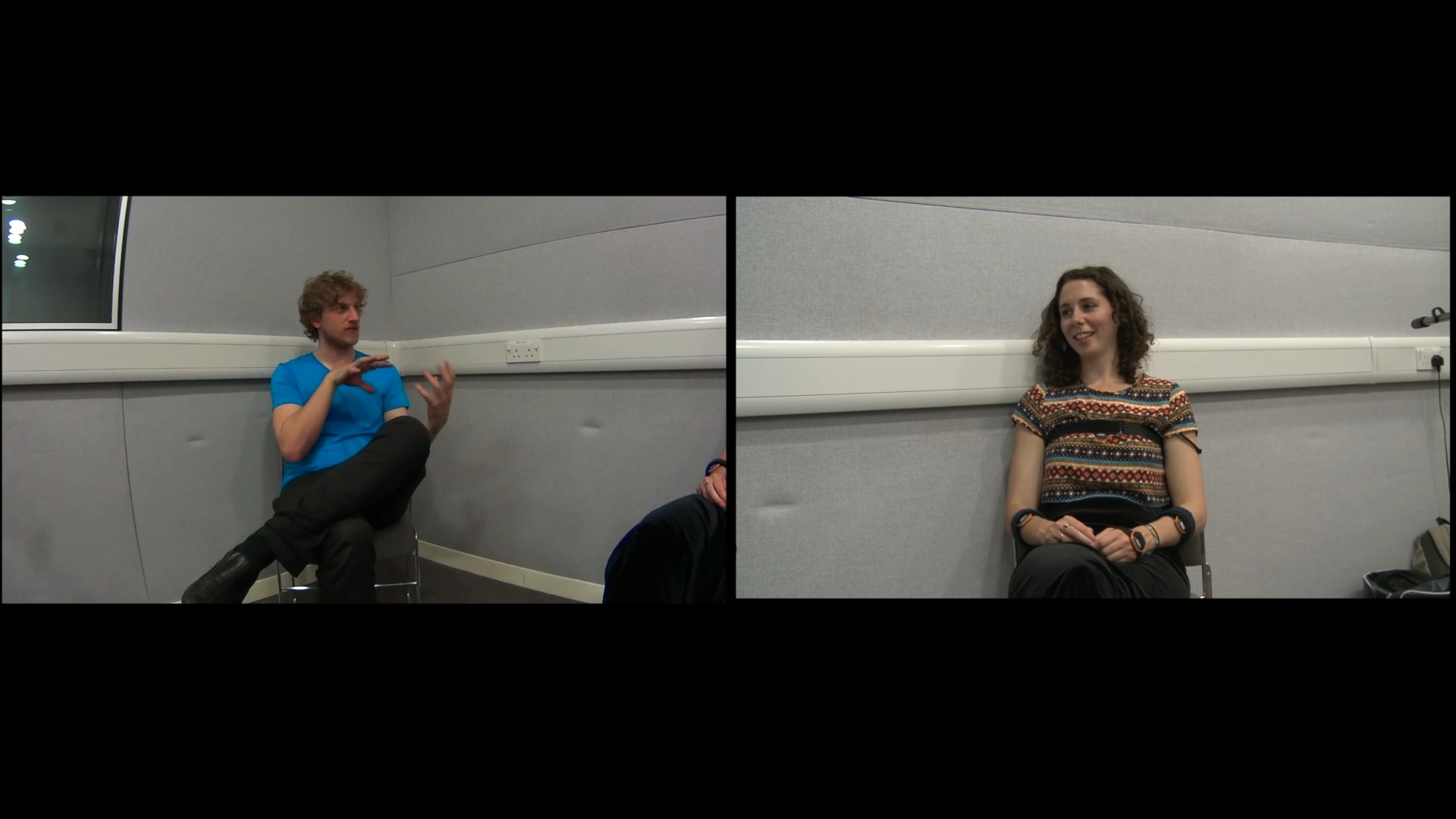}}
\end{minipage}
\caption{Both Speakers During A Dialogue}
\label{fig:screenshot_1}
\end{figure}

\vspace{0.00mm}

\begin{figure}[!htb]
\begin{minipage}{1.0\linewidth}
  \centering
  \centerline{\includegraphics[width=7.5cm]{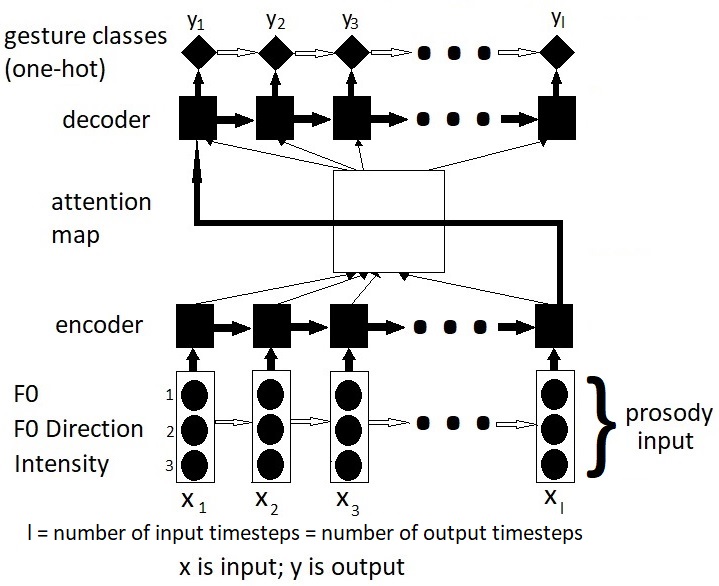}}
\end{minipage}
\caption{The Neural Network Model}
\label{fig:nn_model}
\end{figure}

\vspace{0.00mm}

The corpus has been annotated along different layers~\cite{saintamand2018gestis}: gesture phases (preparation, pre-stroke hold, stroke, post-stroke hold, partial retraction, retraction, and recoil), gesture types (iconic, metaphoric, concrete deixis, abstract deixis, nomination deixis, beat, and emblems), chunk boundaries, classification annotations on whether the gesture is communicative (i.e. contributing to the dialogue discourse) or non-communicative (e.g. rubbing the eyes or scratching nose), the transcription with timestamps. The gesture annotations only consider gestures which are performed by at least one hand. The transcription timestamps include the starting timestamps and the ending timestamps of each word.

We divide the communicative gestures into beats and ideational gestures (i.e. iconic, metaphoric, etc). As explained above beat gestures appear often during the theme while the other gesture types during the rheme. Theme and rheme are marked by different prosodic features \cite{halliday1973explorations,HIR86}. We also divide the gesture phases into strokes and non-strokes. Strokes are often temporally aligned with pitch accent. Therefore, we classify the gestures into four classes:
\begin{itemize}
    \item ``NoGesture'': when no gesture is done
    \item ``Beat'': when beat gesture is done
    \item ``IdeationalOther'': when a non-stroke phase (e.g. preparation, retraction) of an ideational gesture is done
    \item ``IdeationalStroke'': when the stroke phase of an ideational gesture is done
\end{itemize}

\section{Feature Extraction} \label{sec:feature_extraction}
We decompose the speech into utterances where an utterance is defined by sequence of words surrounded by pauses. One utterance is one sample. To define the utterance boundaries, we use the concept of Inter-Pausal Unit (IPU)~\cite{levitan2011measuring}: two consecutive utterances are separated by a silence of at least 200 milliseconds long~\cite{peshkov2013prosodic}.

We use OpenSmile~\cite{eyben2010opensmile} to extract the audio features with 100 milliseconds time-step. We choose 3 prosody features, fundamental frequency / F\textsubscript{0}, F\textsubscript{0} direction score, and intensity, for their temporal relation with gestures~\cite{loehr2012temporal,cravotta2019effects}. We also extract the Mel-frequency cepstral coefficients (MFCC), which is represented as a 13-dimensional vector. MFCC has been successfully used to generate body movements~\cite{hasegawa2018evaluation,kucherenko2019analyzing}.

We also extract eyebrow movements by using OpenFace~\cite{baltrusaitis2018openface}. There are 3 relevant action units (AU): AU1 (inner brow raiser), AU2 (outer brow raiser), and AU4 (brow lowerer). AU 1 and 2 represent rising eyebrow while AU 4 represents lowering eyebrow.

After we obtain the raw AU values, we filter out those whose confidence value is below 0.85 or the AU is absent. Then, we group them into consecutive blocks and we eliminate those whose average value is less than 1. This is done to eliminate noisy data.

The samples are natural utterances that have different lengths. Thus, we pad the sequences to make them have the same length. We pad the inputs with 0-vectors and we pad the outputs with the ``suffix'' auxiliary class. In our full dataset, we have 4161 time-steps of ``NoGesture''s (6.14\%), 1106 time-steps of ``Beat''s (1.63\%), 4208 time-steps of ``IdeationalOther''s (6.20\%), 2739 time-steps of ``IdeationalStroke''s (4.04\%), and 55616 time-steps of the auxiliary ``suffix''s (81.99\%). In total, we have 798 samples.

\section{Model} \label{sec:model}

We use recurrent neural network with attention mechanism~\cite{bahdanau2014neural} to perform the prediction. We use the model which we propose in our previous work~\cite{yunus2019gesture}.

\subsection{Problem Statement} \label{subsec:problem_statement}
Let $X$ be the input and $Y$ be the output. Both $X$ and $Y$ are sequences with the same length. Onward, we will refer to their length as $l$. $X$ is a sequence of vector. Let $X_{i}$ be the vector at timestep $i$, $X_{i}$ is a 3-dimension vector of real numbers containing the three speech prosody features, namely the fundamental frequency (F\textsubscript{0}), the F\textsubscript{0} direction score, and the intensity. $Y$ is a sequence of gesture class (Formulae \ref{eq:yi-def} and \ref{eq:classes_def}).

\vspace{0.00mm}
\useshortskip

\begin{equation} \label{eq:x_timestep}
  X_{i} = (F\textsubscript{0}, F\textsubscript{0}\ direction\ score, intensity)\in \mathbb{R}^3,
\end{equation}

\vspace{0.00mm}
\useshortskip

\begin{equation} \label{eq:classes_def}
    CLASSES = \{NoGesture, Beat, IdeationalOther, IdeationalStroke, Suffix\}
\end{equation}

\vspace{0.00mm}
\useshortskip

\begin{equation} \label{eq:yi-def}
   Y_i \in CLASSES
\end{equation}

\vspace{0.00mm}
\useshortskip

\subsection{Model Overview and Implementation}
The recurrent neural network with attention mechanism is an extension of the encoder-decoder model. The standard encoder-decoder model compresses the entire information from the input sequence into the last encoder node. The attention mechanism adds an attention map between the encoder and the decoder. The map itself is a neuron matrix of the size $l^2$. If $w_{ij}$ is the weight in the attention map at position $\langle i, j \rangle$, then $w_{ij}$ represents the weight of the input at timestep $i$ on the output at timestep $j$. This neuron matrix enables focusing the ``attention'' toward some specific input timesteps. Because this is a multi-class classification problem, we use a one-hot encoding to encode $Y_{i}$. The model schema is in Figure \ref{fig:nn_model}.

We implement the code by using the Zafarali~\footnote{https://github.com/datalogue/keras-attention}\textquotesingle s code as the template. The code is written in Keras~\footnote{https://keras.io/}. We replace the input of the original code \footnote{Originally for date format translation (e.g. the input is ``Saturday 9 May 2018'' string and the output is ``2018-05-09'' string)} by the input we describe in Sub-Section \ref{subsec:problem_statement}. We use categorical cross-entropy as the loss function and Adam as the optimization method. To deal with the class imbalance, we assign weights inversely proportional to the class frequency.

\section{Evaluation Measure} \label{sec:evaluation_measure}
The prior works which also use encoder-decoder model like us use domain specific measurements to evaluate their model. Sutskever et al~\cite{sutskever2014sequence} use BiLingual Evaluation Understudy (BLEU) to evaluate their language translator. Chorowski et al~\cite{chorowski2015attention} use phoneme error rate (PER) to evaluate their speech recognizer. Meanwhile, Bahdanau et al~\cite{bahdanau2016end} use Character Error Rate (CER) and Word Error Rate (WER) to evaluate their speech recognizer. 
 
\begin{figure}[!htb]
    \centering
    \begin{subfigure}{0.4\linewidth}
	\centering
	\centerline{\includegraphics[width=\linewidth]{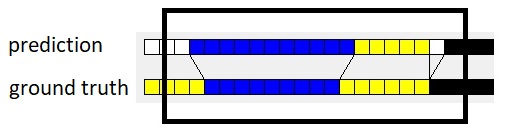}}
	\caption{Alignment}
	\label{fig:alignment_example}
    \end{subfigure}
    \begin{subfigure}{0.4\linewidth}
	\centering
	\centerline{\includegraphics[width=\linewidth]{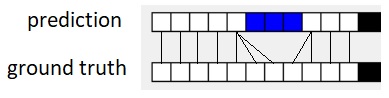}}
	\caption{Discontinuity}
	\label{fig:discontinuity_example}
    \end{subfigure}
    \begin{subfigure}{0.4\linewidth}
	\centering
	\centerline{\includegraphics[width=\linewidth]{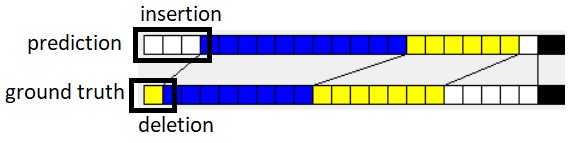}}
	\caption{Insertion and Deletion}
	\label{fig:insertion_deletion_example}
    \end{subfigure}
    \captionsetup{justification=centering}
    \caption{Each cell is 100 ms long. White: ``NoGesture'', Yellow: ``IdeationalOther'', Blue: ``IdeationalStroke''}
\end{figure}

There is not always a gesture on every pitch accent. Moreover gesture stroke may precede the speech prominence. Thus, our evaluation technique should tolerate shifts and dilations to a certain extent. It means that the technique must tolerate that the matching blocks can start at different times and can have different lengths to a certain extent. For example, in Figure \ref{fig:alignment_example}, the predicted ``IdeationalStroke'' starts 100 ms earlier and is 200 ms longer.

Dynamic Time Warping~\cite{bellman1959adaptive} is a sequence comparator which tolerate shifts and dilations. However, this technique does not have a continuity constraint. That is, two consecutive elements which belong to the same class in a sequence might be matched against 2 non-consecutive elements. Without the continuity constraint, we might end up with a match like in Figure \ref{fig:discontinuity_example}. In that figure, we can see that the ``NoGesture''s in the middle of the ground truth are matched with the ``NoGesture''s in the prediction before and after the ``IdeationalStroke''. However, a continuous ``NoGesture'' is different from a ``IdeationalStroke'' preceded and followed by ``NoGesture''.

Thus, we propose a sequence comparison technique to quantify the similarity between the ground truth and the prediction where a block of consecutive elements with the same class is matched against a block of consecutive elements of that class. We use this technique to evaluate our result.

Our measurement uses the sequence comparison algorithm proposed by Dermouche and Pelachaud~\cite{dermouche2016sequence}. It measures the city-block distance between a block in the ground truth and a block in the prediction. This distance metric tolerates shift and dilation up to a certain threshold. If the distance between the 2 blocks is below the threshold, then they are aligned. We define $b_{ps}$ and $b_{pe}$ respectively as the start and the end of the prediction block. Correspondingly, we define $b_{ts}$ and $b_{te}$ respectively as the start and the end of the ground truth block. We also define $T$ as the distance threshold. We define the alignment condition between the prediction block and ground truth block in Formula \ref{eq:match_condition}.

\vspace{0.00mm}
\useshortskip

\begin{equation}
    \label{eq:match_condition}
    ALIGNED \iff \left| b_{ps} - b_{ts} \right| + \left| b_{pe} - b_{te} \right| \leq T
\end{equation}

\vspace{0.00mm}
\useshortskip

We measure the alignment based on how many blocks are aligned and we normalize it against the lengths of those blocks and the frequency of that particular class. Basically, we try to find out for how many time-steps the prediction is aligned to the ground truth, subject to the condition that consecutive time-steps in the ground truth which share the same class must be matched to consecutive or the same time-steps in the prediction which belong to that class as well. This is then normalized against the frequency of that class.

We also introduce the concept of ``insertion'' and ``deletion''. A block which exists in the prediction but has no match in the ground truth is considered to be ``inserted''. This is conceptually similar to \textit{false positive}. The block exists in the prediction but it does not exist in the ground truth.  Similarly, a block which exists in the ground truth but has no match in the prediction is considered to be ``deleted''. This is similar to \textit{false negative}. For example, in Figure \ref{fig:insertion_deletion_example}, we observe an ``inserted'' ``NoGesture'' block and a ``deleted'' ``IdeationalOther'' block. The precise definition of alignment, insertion, and deletion score are at Formulae \ref{eq:del.ins.ali_score}. In the Formulae, $n$ stands for the number of samples in the dataset, $t_c$ is the timestep count of class \textit{c} in the dataset, $p_c$ is proportion of class \textit{c} in the dataset, $l$ is sample length (which is the same for all samples), $b.d$ stands for deleted block, $d_c$ is the deletion score of class \textit{c}, $b.i$ stands for inserted block, $b.p$ stands for predicted block, $b.t$ stands for ground truth block, and $a_c$ is the class \textit{c}\textquotesingle s alignment score. The ideal alignment score is 1 while the ideal deletion and insertion score are 0. It means everything is aligned and there is neither deleted nor inserted block. The insertion score of class \textit{c} can exceed 1 if we predict class \textit{c} more frequently than it actually occurs. On the other hand, the deletion score is always between 0 and 1. The deletion score of class \textit{c} is 1 when we fail to predict any of the block of that class. For the alignment score, if the predictor is accurate but slightly overestimates the length of the block, then the alignment score will be slightly higher than 1. On the other hand, if the predictor is accurate but often slightly underestimates the length of the block, then the alignment score will be slightly lower than 1.

\vspace{0.00mm}
\useshortskip

\begin{equation}
\label{eq:del.ins.ali_score}
\left.\begin{aligned}
p_c &= \frac{t_c}{n  \times l} \\
d_c &= \frac{\Sigma_{b.d} length(b.d)}{n \times l \times p} \\
i_c &= \frac{\Sigma_{b.i} length(b.i)}{n \times l \times p} \\
a_c &= \frac{\Sigma_{(b.p,b.t).aligned} (length(b.p) + length(b.t))}{2 \times n \times l \times p}
\end{aligned}\right.
\end{equation}

\vspace{0.00mm}

\section{Objective Experiments} \label{sec:experiment}
In \textbf{Experiment 1 (random output)}, we generate random outputs only according to the probability distribution of the gesture classes. Specifically, we measure two sets of probabilities, namely the probabilities that a sample is started by a particular class and the probabilities that a class follows another (or the same) class. This is done because our data consist of sequences, where each element affects the next element. We match this result against the output from our ground truth. We do this 55 times and we measure the average of their performances. This can be seen as an extremely simple predictor and thus can be seen as the baseline result.

In \textbf{Experiment 2 (using neural network with the entire dataset)}, we build a neural network model and then we train it. Besides that, We also check whether the validation performance is a reliable proxy of the performance on the testing performance. In a regular machine learning work, we train the model several times, validate each of them, and choose the model with a good performance in the validation. This is based on the assumption that the validation performance is a reliable proxy of the testing performance. We run each of the trained models on both the validation and the testing data set. For each gesture class, we measure the correlation between the alignment scores in the 2 data sets.

In \textbf{Experiment 3 (ablation study)}, we want to observe how different prosody features affect the model performance. We use the model from Experiment 2, but we replace some or all input features (intensity, fundamental frequency / F\textsubscript{0}, and F\textsubscript{0} direction score) with random values. Thus, we render those features useless and we force the model to rely only on the remaining features.

In \textbf{Experiment 4 (inclusion of eyebrow movements)}, we want to find out whether inclusion of eyebrow movements helps on predicting beat class. Eyebrow movements often mark speech prosody and are aligned with pitch accent~\cite{bolinger1989intonation,ekman1979human}. We include the eyebrow movements in the ``Beat'' class. We compare the model performance when the data includes only hand movements, when the data considers hand movements and upward eyebrow movements (Action Unit / AU 1 or 2), and when the data considers hand movements and both upward and downward eyebrow movements (AU 1, 2, or 4). We measure the alignment, insertion, and deletion scores of the ``Beat'' class and we also measure the validation reliability.

In \textbf{Experiment 5 (MFCC as input)}, we use the MFCC instead of prosody as the input features for our neural network. We measure the performance and the validation reliability.

In \textbf{Experiment 6 (both MFCC and prosody as input)}, we use both the MFCC and the prosody as the input features for our neural network. We measure the performance and the validation reliability.

In \textbf{Experiment 7 (trained with one speaker, tested on the other speaker)}, we train the model with one speaker of the dyad in our corpus and test it on the second speaker, and then we do the reverse. It should be noted that one speaker is a man and the other one is a woman.

In Experiments 1, 2, 3, 4, 5, and 6, we partition the full data set into training, validation, and testing data sets identically. We mix all samples from all videos from both speakers and then we randomly split our data with the proportion of 64\%, 16\%, and 20\% for training, validation, and testing data. This is chosen according to the common 80/20 rule. Experiment 7, by its nature, requires us to partition the dataset according to the speaker. We use 80\% of a speaker\textquotesingle s data for training, the remaining 20\% for validation, and 100\% of the data of the other speaker for testing. 

To make the results comparable, we expend equivalent ``effort'' to train the neural network models. We randomly vary the encoder and decoder dimensions from 1 up to the number of features: 3 with prosody, 13 with MFCC, 16 with both prosody and MFCC. We run 25 trainings with 500 epochs, 25 trainings with 1000 epochs, and 5 training with 2000 epochs. Therefore, we have 55 models for each problem. To choose the best model during the validation, we use the weighted average of ``Beat Alignment'', ``IdeationalStroke Alignment'', ``IdeationalOther Alignment'', and ``NoGesture Alignment'' scores, subject to the constraint that each of them must be at least 0.05. The weights are based on the frequency of those classes in the data set. A challenge we face is that the loss function used in the training concerns only the matches at the same timestep, therefore ignoring the possibilities of shifts or dilations, which means that the network is not completely optimized for our objective. Therefore, we have to rely on the stochasticity of the neural network. This situation triggers a question on whether the performance we see with the validation data set is a reliable proxy of what we will see when we use the testing data set.

\vspace{0.00mm}

\begin{table}[!htb]
\begin{center}
\captionsetup{justification=centering}
\caption{Subjective experiment questions and results}
\label{tab:subjective_experiment}
\begin{tabular}{|c|c|c|}
  \hline
  \multicolumn{3}{|c|}{Naturalness} \\
  \hline
  \multicolumn{3}{|c|}{How natural are the gestures?} \\
  \multicolumn{3}{|c|}{How smooth are the gestures?} \\
  \multicolumn{3}{|c|}{How appropriate are the gestures?} \\
  \hline
  Random Output Score & Model Output Score & p-value \\
  \hline
  8.565 & 9.796 & $\bm{1.040 \times 10^{-5}}$ \\
  \hline
  \multicolumn{3}{|c|}{Time Consistency} \\
  \hline
  \multicolumn{3}{|c|}{How well does the gesture timing match the speech?} \\
  \multicolumn{3}{|c|}{How well does the gesture speed match the speech?} \\
  \multicolumn{3}{|c|}{How well does the gesture pace match the speech?} \\
  \hline
  Random Output Score & Model Output Score & p-value \\
  \hline
  8.565 & 10.409 & $\bm{7.271 \times 10^{-9}}$ \\
  \hline
  \multicolumn{3}{|c|}{Semantic Consistency} \\
  \hline
  \multicolumn{3}{|c|}{How well do the gestures match the speech content?} \\
  \multicolumn{3}{|c|}{How well do the gestures describe the speech content?} \\
  \multicolumn{3}{|c|}{How much do the gestures help you understanding the speech content?} \\
  \hline
  Random Output Score & Model Output Score & p-value \\
  \hline
  7.855 & 9.457 & $\bm{4.487 \times 10^{-6}}$ \\
  \hline
\end{tabular}
\end{center}
\end{table}

\vspace{0.00mm}

\begin{table}[!htbp]
\captionsetup{justification=centering}
\caption{Alignment: Exists in both prediction and ground truth \\
Insertion: Exists in the prediction only \\
Deletion: Exists in the ground truth only}
\label{tab:result_scores}
\parbox{.45\linewidth}{
\centering
\begin{tabular}{|c|c|c|c|}
  \hline
  \multicolumn{4}{|c|}{Exp 1: Random output result} \\
  \hline
  & Alignment & Insertion & Deletion \\
  \hline
  Beat & \textbf{0.009} & 0.936 & 0.990 \\
  IdeationalStroke & 0.084 & 0.485 & 0.904 \\
  IdeationalOther & 0.109 & 0.563 & 0.882 \\
  NoGesture & \textbf{0.533} & 0.940 & 0.453 \\
  \hline
  \multicolumn{4}{|c|}{Exp 2: Using neural network} \\
  \multicolumn{4}{|c|}{with the entire dataset} \\
  \hline
  Beat & \textbf{0.194} & 3.127 & 0.802 \\
  IdeationalStroke & \textbf{0.507} & 0.485 & 0.582 \\
  IdeationalOther & \textbf{0.304} & 0.226 & 0.671 \\
  NoGesture & 0.567 & 0.554 & 0.398 \\
  \hline
  \multicolumn{4}{|c|}{Exp 4: inclusion of eyebrow movements} \\
  \hline
  & Alignment & Insertion & Deletion \\
  \hline
  Hand Only & \textbf{0.194} & 3.127 & 0.802 \\
  \hline
  With Upward & 0.136 & 1.038 & 0.829 \\
    Eyebrow & & &  \\
    Movement & & &  \\
  \hline
  With Upward \& & \textbf{0.222} & 0.280 & 0.774 \\
    Downward & & & \\
    Eyebrow & & & \\
    Movement & & & \\
  \hline
  \multicolumn{4}{|c|}{Exp 5: MFCC as input} \\
  \hline
  & Alignment & Insertion & Deletion \\
  \hline
  Beat & 0.171 & 2.619 & 0.849 \\
  IdeationalStroke & \textbf{0.166} & 0.977 & 0.855 \\
  IdeationalOther & 0.362 & 0.538 & 0.652 \\
  NoGesture & 0.440 & 0.789 & 0.551 \\
  \hline  
  \multicolumn{4}{|c|}{Exp 6: MFCC and prosody as input} \\
  \hline
  & Alignment & Insertion & Deletion \\
  \hline
  Beat & 0.000 & 2.429 & 1.000 \\
  IdeationalStroke & 0.388 & 0.790 & 0.640 \\
  IdeationalOther & 0.362 & 0.584 & 0.613 \\
  NoGesture & 0.441 & 0.891 & 0.563 \\
  \hline
  \multicolumn{4}{|c|}{Exp 7: trained with one speaker} \\
  \multicolumn{4}{|c|}{tested on the other} \\
  \hline
  \multicolumn{4}{|c|}{Trained on speaker 1, tested on speaker 2} \\
  \hline
  & Alignment & Insertion & Deletion \\
  \hline
  Beat & 0.015 & 1.049 & 0.982 \\
  IdeationalStroke & \textbf{0.506} & 1.142 & 0.559 \\
  IdeationalOther & 0.367 & 0.359 & 0.575 \\
  NoGesture & 0.517 & 0.441 & 0.459 \\
  \hline
  \multicolumn{4}{|c|}{Trained on speaker 2, tested on speaker 1} \\
  \hline
  & Alignment & Insertion & Deletion \\
  \hline
  Beat & 0.132 & 3.679 & 0.856 \\
  IdeationalStroke & \textbf{0.396} & 0.846 & 0.650 \\
  IdeationalOther & 0.217 & 0.221 & 0.746 \\
  NoGesture & 0.538 & 0.589 & 0.424 \\  
  \hline
\end{tabular}
}
\hfill
\parbox{.45\linewidth}{
\centering
\captionsetup{justification=centering}
\begin{tabular}{|c|c|c|c|}
  \hline
  \multicolumn{4}{|c|}{Exp 3: Ablation study} \\
  \hline
  \multicolumn{4}{|c|}{All features are randomized} \\
  \hline
  & Alignment & Insertion & Deletion \\
  \hline
  Beat & 0.040 & 0.643 & 0.929 \\
  IdeationalStroke & 0.038 & 0.072 & 0.952 \\
  IdeationalOther & 0.025 & 0.027 & 0.960 \\
  NoGesture & 0.347 & 0.275 & 0.641 \\
  \hline
  \multicolumn{4}{|c|}{Using intensity only} \\
  \hline
  & Alignment & Insertion & Deletion \\
  \hline
  Beat & 0.0 & 0.786 & 1.000 \\
  IdeationalStroke & 0.077 & 0.063 & 0.922 \\
  IdeationalOther & 0.039 & 0.040 & 0.936 \\
  NoGesture & 0.376 & 0.298 & 0.589 \\
  \hline
  \multicolumn{4}{|c|}{Using F\textsubscript{0} and the F\textsubscript{0} direction score only} \\
  \hline
  & Alignment & Insertion & Deletion \\
  \hline
  Beat & \textbf{0.175} & 2.444 & 0.802 \\
  IdeationalStroke & \textbf{0.481} & 0.503 & 0.563 \\
  IdeationalOther & \textbf{0.313} & 0.179 & 0.637 \\
  NoGesture & \textbf{0.596} & 0.555 & 0.379 \\
  \hline
  \multicolumn{4}{|c|}{Using F\textsubscript{0} only} \\
  \hline
  & Alignment & Insertion & Deletion \\
  \hline
  Beat & \textbf{0.179} & 2.540 & 0.802 \\
  IdeationalStroke & \textbf{0.521} & 0.515 & 0.553 \\
  IdeationalOther & \textbf{0.273} & 0.155 & 0.664 \\
  NoGesture & \textbf{0.577} & 0.570 & 0.393 \\
  \hline
  \multicolumn{4}{|c|}{Using F\textsubscript{0} direction score only} \\
  \hline
  & Alignment & Insertion & Deletion
  \\
  \hline
  Beat & 0.044 & 0.548 & 0.929 \\
  IdeationalStroke & 0.024 & 0.083 & 0.965 \\
  IdeationalOther & 0.019 & 0.013 & 0.969 \\
  NoGesture & 0.379 & 0.311 & 0.630  \\
  \hline
\end{tabular}
}
\end{table}

\vspace{0.00mm}

\begin{table}[!htbp]
\begin{center}
\captionsetup{justification=centering}
\caption{Validation reliability}
\label{tab:validation_reliabilities}
\begin{tabular}{|c|c|c|c|}
  \hline
  \multicolumn{4}{|c|}{Exp 2: Using neural network with the entire dataset} \\
  \hline
  Alignment Score of ... & Mean at Validation Data & Mean at Testing Data & Correlation \\
  \hline
  Beat & \textbf{0.202} & \textbf{0.244} & \textbf{-0.037} \\
  IdeationalStroke & 0.317 & 0.361 & 0.875 \\
  IdeationalOther & 0.202 & 0.274 & 0.809 \\
  NoGesture & 0.537 & 0.546 & 0.679  \\
  \hline
  \multicolumn{4}{|c|}{Exp 4: inclusion of eyebrow movements, ``Beat'' alignment score} \\
  \hline
  Alignment Score of ... & Mean at Validation Data & Mean at Testing Data & Correlation \\
  \hline
  Hand Only & 0.202 & 0.2444 & -0.037 \\
     & & & \\
  \hline
  With Upward & 0.078 & 0.102 & \textbf{0.414} \\
    Eyebrow Movement & & &  \\
  \hline
  With Upward/Downward & 0.226 & 0.219 & \textbf{0.925} \\
    Eyebrow Movement & & & \\
  \hline
  \multicolumn{4}{|c|}{Exp 5: MFCC as input} \\
  \hline
  Alignment Score of ... & Mean at Validation Data & Mean at Testing Data & Correlation \\
  \hline
  Beat & 0.060 & 0.084 & \textbf{-0.056} \\
  IdeationalStroke & 0.248 & 0.256 & 0.405 \\
  IdeationalOther & 0.283 & 0.340 & 0.502 \\
  NoGesture & 0.452 & 0.467 & 0.204  \\
  \hline
  \multicolumn{4}{|c|}{Exp 6: both MFCC and prosody as input} \\
  \hline
  Alignment Score of ... & Mean at Validation Data & Mean at Testing Data & Correlation \\
  \hline
  Beat & 0.080 & 0.0745 & \textbf{0.025} \\
  IdeationalStroke & 0.272 & 0.265 & 0.472 \\
  IdeationalOther & 0.302 & 0.351 & 0.622 \\
  NoGesture & 0.425 & 0.465 & 0.386  \\
  \hline
\end{tabular}
\end{center}
\end{table}

\section{Subjective Experiment} \label{sec:subjective_experiment}
In the subjective experiment, 31 respondents watched 12 videos online of a virtual agent speaking and performing communicative gestures. Among them, 17 (55\%) are male, 13 (42\%) are female, and 1 (3\%) refuses to disclose the gender. On the age breakdown, 6 (19\%) are 18-20 years old, 20 (65\%) are 21-30 years old, 2 (6\%) are 31-40 years old, and 3 (10\%) are 41-50 years old.

The 12 videos consist of 6 pairs. We extract 6 segments from the Gest-it corpus (3 segments with a man, 3 segments with a woman). We replicate the real human gestures on the virtual agent. We match the human gender to the agent gender. Each pair of videos consists of the baseline and the gesture generation model\textquotesingle s output. The gesture timing of the baseline videos is decided by randomly shuffling the timing from the ground truth. In both baseline and model output videos, we retain the ground truth\textquotesingle s gesture shapes. In both videos, the agents have the same appearance and say the same sentence. We also use the original voice from the corpus. Thus, the differences in the video pairs are only in the gesture timings. The agent animation contains only the arm gestures. There is no other animation (no head motion, gaze, posture shift, etc). Moreover, we blur the face of the agent because its still blank face could have distracted the respondents. The sequence of the 12 videos is shuffled so that a pair will not be shown consecutively. 

Our objective is to compare the respondent\textquotesingle s perception differences between the videos based on the model output and the baseline videos. We compare the naturalness, the time consistency, and the semantic consistency of the videos. For each of those dimensions, we measure it by asking the respondents to answer 3 questions. Each question asks the user to give a rating in likert scale from 1 to 5. We sum the respondent\textquotesingle s scores on the three questions to get the score of that dimension. We adapt the questions from the subjective study of Kucherenko et al~\cite{kucherenko2019analyzing}. We find that in all the 3 dimensions, the videos created based on the model output have higher average score. We also check the significances by using one-way ANOVA test. The questions and results are in Table \ref{tab:subjective_experiment}.

\section{Discussion} \label{sec:discussion_and_conclusion}
We observe in ``Exp 1: Random output result'' (Table \ref{tab:result_scores}), different classes have different complexities. For example, ``Beat'' classifier (alignment score = 0.009) needs a higher Vapnik-Chervonenkis (VC) dimension than the classifiers of another classes do.  VC dimension is an abstract measure of how complex a classifier function can be. Meanwhile, the ``NoGesture'' class, with (alignment score = 0.533), can work with a lower VC-dimensioned classifier, despite the fact that we select our samples only when the person is speaking.

Experiment 1 result might be caused by the data imbalance. The ``NoGesture'' class is almost 300\% larger than the ``Beat'' class. The ``Beat'' rarity might cause the prediction to have a lower performance. Besides that, our corpus is small (798 samples), which makes the training hard.

When we run our network (``Exp 2: Using neural network with the entire dataset'', Table \ref{tab:result_scores}), we observe that the alignment scores outperform the random output on all classes. It suggests that the 3 prosody features (F\textsubscript{0}, F\textsubscript{0} direction score, and intensity) enable prediction of the gesture classes with a certain degree of reliability. However, the ``Beat'' class result is not reliable. As we observe in ``Exp 2: Using neural network with the entire dataset, Validation reliability'' (Table \ref{tab:validation_reliabilities}), the correlation between the validation performance and the testing performance is almost zero. It means that in respect to the ``Beat'' class, validation is useless, thus the testing result can be attributed to chance. However, the mean alignment scores of the ``Beat'' class is still higher than the random output (``Exp 1: Random output result'', Table \ref{tab:result_scores}), which suggests that the neural network still learns some pattern. However, as we have noted earlier, ``Beat'' is rare in our corpus.

Therefore, we wonder if we can predict ``Beat'' better should we have more data. Besides that, ``Beat'' gestures can also be performed by head or facial movements~\cite{bolinger1989intonation,ekman1979human,krahmer2004more}. Indeed, in Experiment 4 we find the ``Beat'' class alignment score is slightly higher when we include both the upward and downward eyebrow movements (``Exp 4: inclusion of eyebrow movements'', Table \ref{tab:result_scores}). More importantly, the validation reliability markedly improves (``Exp 4: inclusion of eyebrow movements'', Table \ref{tab:validation_reliabilities}).  These results shows that beat gestures can indeed be performed by eyebrow movements. Therefore, including eyebrow movements increases the amount of ``Beat'' data and, thus, enhances our model\textquotesingle s reliability.

On the ``IdeationalStroke'' class, our predictor surpasses the random output generator. This class encompasses the stroke of all communicative gestures except beat gestures. The model can predict where a gesture stroke is aligned with the acoustic features. This phase is well-studied in gesture literature as it carries the gesture meaning. This phase usually happens around or slightly before the pitch accent~\cite{wagner2014gesture}. In our case, we have the intensity, F\textsubscript{0}, and F\textsubscript{0} direction score as our input. They participate to the characterization of the pitch accent. We also find that our result is reliable, because the alignment scores at the validation data set and at the testing data set show a positive correlation. 

On the ``IdeationalOther'' class the model yields an alignment score higher than the random output, but the alignment score is still low. As a recall, this class contains all the gesture phases (e.g., preparation, hold, retraction) except the stroke phase for all ideational gestures. We can notice that, in all our experiments, we never obtain a good alignment on this class. This class is made of different gesture phases that may not correspond to the same prosodic profile. Their alignment may obey to different synchronisation needs \cite{wagner2014gesture}. However, we still find that our validation result is reliable.

In the ablation study (Experiment 3), we replace some features with random values to observe how it affects the model performance. We start by replacing the entire input with random values and use it on the trained model (``Exp 3: Ablation study, All features are randomized'', Table \ref{tab:result_scores}), we observe that all the alignment scores are lower than in the random output result, except for ``Beat'' which is 0.040, which only marginally outperforms the random output. Subsequently, when we use the intensity alone (``Exp 3: Ablation study, Using intensity only'', Table \ref{tab:result_scores}), we find again that the model\textquotesingle s alignment scores fail to outperform the random output result. This result does not prove either that it is impossible to learn the gesture timing from the intensity. Our model simply happens to largely ignore the intensity feature, yet it still can predict some classes (as shown in Experiment 2 results). Finally, in the sub-experiment where we only use fundamental frequency ``Exp 3: Ablation study, Using F\textsubscript{0} only'', Table \ref{tab:result_scores}), the alignment scores are similar to what we get when we use all prosody features (``Exp 2: Using neural network with the entire dataset'', Table \ref{tab:result_scores}). This result suggests that F\textsubscript{0} is tied and is very pertinent to the gesture timing.

In Experiment 5 where we use MFCC instead of the prosody features (``Exp 5: MFCC as input'', Table \ref{tab:result_scores}), we find that the alignment scores of ``Beat'', ``IdeationalStroke'', and ``IdeationalOther'' outperform the random output. However, the ``IdeationalStroke'' alignment score is considerably lower than when we use prosody features (``Exp 2: Using neural network with the entire dataset'', Table \ref{tab:result_scores}). A possible reason is because the MFCC are represented as a 13-dimensional vector while the prosody features are represented as a 3-dimensional vector. The higher dimension makes the search space much larger, and thus making the training slower. Another possible reason is that the MFCC is indeed less informative about stroke timing. Indeed, it has been reported in several studies that F\textsubscript{0}/pitch are related to gesture stroke timing~\cite{wagner2014gesture,lucking2016finding}. On the validation reliability, we also find that the correlation between the validation alignment scores and on the testing alignment scores of the ``Beat'' class is close to zero (``Exp 5: MFCC as input'', Table \ref{tab:validation_reliabilities}). This is similar to we when use prosody features (``Exp 2: Using neural network with the entire dataset, Validation reliability'', Table \ref{tab:validation_reliabilities}). 

In Experiment 6 where we use both the prosody features and MFCC (``Exp 6: both MFCC and prosody as input'', Table \ref{tab:result_scores}), the alignment score of the ``Beat'' falls to 0.000 while the alignment score of ``IdeationalOther'' increases to 0.362. The alignment score of ``Beat'', like in Experiments 2 and 5, can be attributed to chance as shown by the almost zero correlation in the validation reliability test (``Exp 6: both MFCC and prosody as input'', Table \ref{tab:validation_reliabilities}). The alignment score of ``IdeationalOther'', which is still lower than what we get when we use prosody features only, can likely be contributed to the presence of the prosody features, especially the F\textsubscript{0} which we have shown to be pertinent to gesture timing. Although having more features enables the neural network to learn more information, it also makes the search space larger, which in turn makes the search slower.

In Experiment 7 where we train the model with one speaker and test it on the other speaker of the same interaction (``Exp 7: trained with one speaker tested on the other'', Table \ref{tab:result_scores}), we find that the models\textquotesingle alignment scores outperform the random output, which suggests that some generalizability exists even-though people have different gesturing styles. These results may also be due as both speakers are part of the same interaction and conversation participants tend to automatically align to each other, at different levels, such as phonology, syntax and semantics~\cite{menenti2012toward}, as well as gesture types~\cite{wessler2017temporal}. These different alignments make the conversation itself successful~\cite{garrod2009joint}.

In our subjective experiment, we measure the naturalness, time consistency, and semantic consistency of the gestures and speech. We compare the perception by human participants of the animation of the virtual agent where we manipulated the timing of the gestures. It allows measurement of the impact of the timing generated by the neural network against random timing along the 3 qualities: naturalness of the agent gesturing, time consistency of the gesture production and of the speech prosody, and the semantic alignment of both. The random timing acts as the baseline. The idea is similar to what we do in our objective evaluations (Experiments 1 and 2). We find that the timing from the model outperforms the baseline in all measured qualities, and the differences are significant ($p-value < 0.05$). It shows that overall the generated result is perceived better by the human respondents along the 3 qualities. It also shows that  gesture timing is important to how well-perceived the gestures are by humans. We keep the gesture shapes from the ground truth in both the output of our model and the baseline, we act only on the timing of the gestures, yet the output of our model is perceived more favourably.

\section{Conclusion and Future Work} \label{sec:future_work}
In this paper we have presented a model to predict where to place gestures based only on the acoustic features. We limit the scope of the problem to only the gesture timing. We use 3 prosodic features as the input, namely the fundamental frequency (F\textsubscript{0}), F\textsubscript{0}) direction score, and intensity. We also experiment with using Mel-frequency cepstral coefficients (MFCC) as the input. We consider 2 classes of communicative gestures (beats and ideational gestures) and 2 classes of gesture phases (stroke and others). In an experiment we also add eyebrow movements that can have communicative functions, such as being prosodic markers. We conduct several objective studies to evaluate the model as well as a subjective study. Our results show the pertinence of the F\textsubscript{0} to determine gesture timing. We also find that considering eyebrow movements as beat gestures increases the beat prediction accuracy.

However, gesture generation is also tightly linked to what is being said. In the future, our aim is to consider not only the prosody but also the semantics of the speech. The question of representing the semantics arises. We are planning to rely on a higher representation level such as image schema that can be linked to metaphoric gestures~\cite{johnson2013body}. Combining both of semantics and the prosody is a challenge by itself. A big part of the challenge is that aligning meaning, prosody and gestures is far from being a trivial problem. In the future, we intend to go into this direction.

\section*{Acknowledgement}
This project has received funding from the European Union\textquotesingle s Horizon 2020 research and innovation programme under grant agreement No 769553. This result only reflects the authors\textquotesingle \space views and the European Commission is not responsible for any use that may be made of the information it contains. 
We thank Katya Saint-Amand for providing the Gest-IS corpus~\cite{saintamand2018gestis}.

\bibliographystyle{acm}
\bibliography{mypaper}

\end{document}